\documentclass[12pt,leqno]{article}
\usepackage{amsmath,amsfonts}
\usepackage{graphicx}
\usepackage{amsthm,amssymb,bm}
\usepackage{setspace}

\begin{document}
\title{Geometrical Theory of Separation of Variables, a review of recent developments}
\author{Giovanni Rastelli \\ \\ Dipartimento di Matematica, Universit\`a di Torino. \\ Torino, via Carlo Alberto 10, Italia.\\ \\ e-mail: giorast.giorast@alice.it }
\maketitle

\bigskip

\begin{abstract}
The Separation of Variables theory for the Hamilton-Jacobi equation is 'by definition' related to the use of special kinds of coordinates, for example Jacobi coordinates on the ellipsoid  \footnote{C. G. Jacobi, {\it Vorlesunger \"uber Dynamik}, Gesammelte Werke, Berlin (1884)}  or St\"ackel systems \footnote{P. St\"ackel, Math. Ann. {\bf42} 537 (1893)}  in the Euclidean space. However, it is possible and useful to develop this theory in a coordinate-independent way: this is the
Geometrical Theory of Separation of Variables. It involves geometrical objects (like special submanifolds and foliations) as well as special vector and tensor fields like Killing vectors and Killing two-tensors (i.e. isometries of order one and two), and their conformal extensions; quadratic first integrals are associated with the Killing two-tensors. In the recent years Separable Systems provide mathematical structures studied from different points of view. We present here a short review of some of these structures and of their applications with particular consideration to the underlying geometry. Algebraic relations among Killing tensors, quadratic first integrals or their associated second order differential operators and some aspects of approximation with separable systems are considered. This paper has been presented as a poster at Dynamics Days Europe 2008, Delft 25-29 August 2008.
\end{abstract}

\section*{Introduction} 
There are  two levels of the Geometrical Theory of Separation of Variables (GTSOV): (i) general or non-orthogonal separation (i.e., not necessarily orthogonal), (ii) orthogonal separation. A geometrical characterization (in terms of foliations and Killing vectors and tensors) of the general separation has been proposed by Benenti \footnote{S. Benenti, J.Math.Phys. {\bf 38} 6578 (1997)} . Among the applications of such a characterization we find a coordinate-independent proof of a theorem of Kalnins and Miller \footnote{E. Kalnins and W. Miller Jr, SIAM J. Math. Anal. {\bf 11} 1011 (1980)} :  on a Riemannian manifold with constant curvature the SOV always occurs in orthogonal coordinates. A second application is a finer classification of the orthogonal separable systems. Eisenhart's theorem on St\"ackel systems \footnote{L. P. Eisenhart, {\it Riemannian geometry}, Princeton University Press (1949)}  provides a first result on the geometrical characterization of the Orthogonal-SOV (OSOV).
However, the necessary and sufficient conditions written in the Eisenhart statements are redundant. Minimal conditions have been proposed by Benenti. According to one of the various possibilities, OSOV of the geodesic flow on a Riemannian manifold occurs if and only if there exists a Killing tensor $\mathbf K$ with simple eigenvalues and normal eigenvectors. Such a tensor has been called a {\bf characteristic tensor}. 
Furthermore, if a potential energy $V$ is present, it must satisfy the {\bf characteristic equation} $d\mathbf K\;dV=0$. With respect to Cartesian coordinates on an Euclidean space this equation reduces to the so-called {\bf Bertrand-Darboux equation}. The existence of a characteristic KT  implies the existence of other $n-1$ linearly independent KT with common eigenvectors $K_i$ which are associated with {\bf $n-1$ quadratic first integrals} of $H$  by $H_i=\frac 12 K^{jl}_ip_jp_l+V_i$ with $dV_i=K_i\;dV$. In many cases  (for example when the Ricci tensor is null) it is possible to build {\bf quantum symmetry operators} for the Schr\"odinger equation associated with $H$ by putting ${\hat H}_i=\nabla_j(K^{jl}_i\nabla_l)+V_i$ and obtaining  multiplicative separation for the Schr\"odinger equation in the same coordinates as for the classical system \footnote{S.Benenti, C.Chanu and G. Rastelli, J.Math. Phys. {\bf 43} 5183 (2002)}. In the following, some few topics of GTSOV are sketched, together with some of their possible applications.

%\footnotemark[2] \footnotetext{P. St\"ackel, Math. Ann. {\bf42} 537 (1893)}

%\footnotemark[3]\footnotetext{S. Benenti, J.Math.Phys. {\bf 38} 6578 (1997)}

%\footnotemark[4]\footnotetext{E. Kalnins and W. Miller Jr, SIAM J. Math. Anal. {\bf 11} 1011 (1980)}

%\footnotemark[5]\footnotetext{L. P. Eisenhart, {\it Riemannian geometry}, Princeton University Press (1949)}

%\footnotemark[6]\footnotetext{S.Benenti, C.Chanu and G. Rastelli, J.Math. Phys. {\bf 43} 5183 (2002)}

\section{Benenti systems} Among the orthogonal separable systems we find the special class of the so-called {\bf Benenti systems} (or {\bf L-systems}) \footnote{S.Benenti, Acta Applicanda Mathematica {\bf 87} 33 (2005)}, for which the separation is characterized by a  special conformal Killing tensor $\mathbf L$ called {\bf Benenti tensor} (or {\bf L-tensor}). For these systems a complete set of quadratic first integrals  can be constructed by a pure algebraic and coordinate independent method starting from $\mathbf L$.
The geodesic flow on an asymmetric ellipsoid (Jacobi) is an example of a Benenti system.
Several recent papers have shown that Benenti systems have a very rich structure closely related to other fields of research. An {\bf L-tensor} is a conformal KT of order (1,1) which is torsionless with (real) simple eigenvalues.
Let L a symmetric 2-tensor, then the tensors $K_a$, $a=0,\ldots ,n-1$, defined by
$$
K_0=I, \quad K_a=\frac 1a tr(K_{a-1}L)I-K_{a-1}L, \quad a>1,
$$
are n independent KT with common normal eigenvectors iif L is a L-tensor.
The KT's $K_a$  determine an orthogonal separable system, a St\"ackel system called Benenti system. Not all the St\"ackel systems are Benenti systems. The first integrals are obtained from the KT as exposed in the Introduction; a recursive formula similar to the previous one can produce the potentials in each of the first integrals. 

%\footnotemark[7]\footnotetext{BS}{\small{S.Benenti, Acta Applicanda Mathematica {\bf 87}} 33 (2005)}

\subsection{Separable coordinates for triangular Newton equations } (see \footnote{K.Marciniak and S. Rauch-Wojciechowski, Studies in Applied Mathematics {\bf 118} 45 (2007)}) Two dynamical systems on the same configuration manifold Q are {\bf equivalent} if their motions on Q locally coincide as unparametrized curves; i.e. up to time transformations of the kind $dt=\mu(q)dt'$ (\footnotemark[7] ). For natural systems: a dynamical system admits a Lagrangian equivalent system iif the force F is such that $F=-A^{-1}\nabla V$, where $A$ is the cofactor of a special conformal Killing tensor $J$: $A=(det J )J^{-1}$. A special conformal KT with simple eigenvalues is a L tensor. An L tensor is a special conformal KT. A cofactor pair system is a cofactor system in two distinct ways.A cofactor pair system such that $\bar J$ has real simple eigenvalues is equivalent to an L system generated by $\bar J$. {\bf The triangular Newton equations} 
$$
\ddot{q_i}=M_i(q_1, \ldots, q_i) \qquad i=1\ldots n
$$
are not in general a Lagrangian system, but if $M=-(cof \; G)^{-1}\nabla k$, where $G$ is the matrix associated with the characteristic KT of elliptic coordinates in the Euclidean space, it is equivalent to a Lagrangian one. In this case in fact $G$ is an L tensor, then a special conformal KT and generates a Benenti system. The system equivalent to the TNE is then  Hamiltonian and admits SOV in the orthogonal coordinates of the Benenti system.

Example :

Let $(x_1,x_2)$ Cartesian coordinates, let the triangular system be $\ddot{x}_1=-4x_1$, $\ddot{x}_2=6x_1^2-4x_2$; this is a cofactor system with $G=\left( \begin{matrix} 1 & -x_1 \cr -x_1 & -2x_2 \end{matrix}\right)$. The separable coordinates (not a St\"ackel system of the Euclidean plane) are $u_1=x_1$ and $u_2=\frac 12 x_1^2+x_2$ and their associated quadratic first integrals provide the quadrature of the system. 

\subsection{Separation curves, Stackel systems and soliton systems }  (see \footnote{M.B\l aszak and K. Marciniak, JMP {\bf 47} 032904 (2006)}) A set of n relations of the form
$\phi_i(\lambda_i,\mu_i, a_1, \ldots, a_n)=0 \qquad a_i\in \mathbb R$,
such that $(\lambda_i,\mu_i)$ are canonical coordinates and $\det(\partial \phi_i/\partial a_j)\neq 0$ are called separation relations, when they are all of the same form are called {\bf separation curve}. If the separation curve is of the form
$$
H_1\lambda^{n-1}+H_2\lambda^{n-2}+\ldots +H_n=\frac 12 \lambda^m\mu^2+\lambda^k
$$
where the $H_i$ are the polynomials $H_i=\frac 12\mu^TK_iG^{(m)}\mu+V^{(k)}_i$ with $G^{(m)}=L^mG^{(0)}$, $G^{(0)}=diag\; (1/\Delta_1,\ldots,1/\Delta_n)$, $\Delta_i=\prod_{j\neq i}(\lambda_i-\lambda_j)$ and where
$L=diag\; (\lambda_1,\ldots,\lambda_n)$ is a conformal KT with respect to the metrics $G^{(m)}$. The tensors $K_i$ are given by 
$K_{i+1}=LK_i+q_iI, \quad K_1=I, K_{n+1}=0$
where the $q_i(\lambda)$ are coefficients of the characteristic polynomial of $L$. The potentials $V_i^k$ are obtained by the recursion $V_i^k=V^{k-1}_{i+1}-q_iV^{k-1}_1$. Therefore, the $H_i$ form a Benenti system generated by the L-tensor $L$ with separable coordinates $\lambda_i$. The $q_i$ are new coordinates on the configuration manifold and  for the Benenti systems of above a hierarchy of integrable dispersionless equations, called Killing dispersionless system, whose solutions are called in general {\bf solitons}, is obtained  from the equation  $\frac d{dt_i}q_j=K_i(q)\frac d{dx}q$
where $t_i$ is the evolution parameter (time) of the Hamiltonian flow of $H_i$ and $x=t_1$.

Example: 

$H_1\lambda+H_2=\frac 12\lambda \mu^2+\lambda^4$
be the separation curve, the diaganalized Hamiltonians (Benenti system) are $H_1=\frac1{2(\lambda_1-\lambda_2)}(\lambda_1\mu_1^2-\lambda_2\mu_2^2+2(\lambda_1^4-\lambda_2^4)$, $H_2=\frac{\lambda_1\lambda_2}{2(\lambda_2-\lambda_1)}(\mu_1^2-\mu_2^2+2(\lambda_1^3-\lambda_2^3)$. $q_1=\lambda_1+\lambda_2$, $q_2^2=-4(\lambda_1\lambda_2)$
where $(\lambda_i,\mu^i)$ are canonical Parabolic coordinates. In Cartesian canonical coordinates $(q_i,p^i)$ they are
$$
H_1=\frac 12(p_1^2+p_2^2)+q_1^3+\frac12 q_1q_2^2, \quad H_2=\frac 12 (q_2p_1p_2-q_1p_2^2+\frac 18q_2^4+\frac 12 q_1^2q_2^2),
$$
$H_1$ is the Hamiltonian of a separable  case of the Henon-Heiles system. The integrable dispersionless PDE are
$$
q_{1,t}=\frac 12 q_2q_{2,x}, \qquad q_{2,t}=\frac 12 q_2q_{1,x}-q_1q_{2,x}
$$
that, after differentiation and elimination of $q_2$ yield the {\bf KdV soliton equation}
$q_{1,t}+\frac 12q_{1,xxx}+3q_1q_{1,x}=0$ together with one differential consequence of it. 

%\footnotemark[8] \footnotetext{MaRa}{\small{K.Marciniak and S. Rauch-Wojciechowski, Studies in Applied Mathematics {\bf 118} 45 (2007)}}\hspace{1cm}
%\footnotemark[9] \footnotetext{BM}{\small{M.B\l aszak and K. Marciniak, JMP {\bf 47} 032904 (2006)}}

\section{Superintegrability} Classical  Hamiltonian systems in $n$ dimensional manifolds admit at most $2n-1$ functionally independent (f.i.) first integrals. They are Liouville integrable if $n$ f.i. first integrals are in involution. Integrable systems with more than $n$ f.i. first integrals are called {\bf superintegrable}. Well known examples are the harmonic oscillator and the Kepler-Coulomb system. Because the integral curves of the systems stay in the intersection of the level hypersurfaces of the first integrals,  the occurrence of the maximum number of functionally independent first integrals, i.e  $2n-1$, implies that the integral curves of the system can be determined by algebraic methods. Hamiltonian systems separable in multiple coordinate systems admits often more than $n$ f.i. quadratic first integrals and they are said to be {\bf quadratically superintegrable}. The Kepler-Coulomb system is an example of quadratically maximally superintegrable system. By searching among multiseparable systems many instances of superintegrability have been found recently and their classification is in progress\footnote{E. G. Kalnins, G. Williams, W. Miller Jr  and G. S. Pogosyan, J. Math. Phys. {\bf 40} 708 (1999)} \footnote{S.Gravel and P.Winternitz, J. Math. Phys. {\bf 43} 5902 (2002)} \footnote{C.Daskaloyannis and K.Ypsilantis, J. Math. Phys. {\bf 47} 042904 (2006)} \footnote{S. Benenti,C. Chanu and G. Rastelli, J. Math. Phys. {\bf 41}, 4654 (2000)} \footnote{P.Tempesta et al. eds: {\it Superintegrability in Classical and Quantum Systems}, CRM proceedings \& Lecture Notes {\bf 37} Amer. Math. Soc. (2004)}.

% \footnotemark[10]\footnotetext{KMK}{E. G. Kalnins, G. Williams, W. Miller Jr  and G. S. Pogosyan, J. Math. Phys. {\bf 40} 708 (1999) }\hspace{1cm}\footnotemark[11]\footnotetext{Win}{S.Gravel and P.Winternitz, J. Math. Phys. {\bf 43} 5902 (2002)}\hspace{1cm}\footnotemark[12]\footnotetext{D}{C.Daskaloyannis and K.Ypsilantis, J. Math. Phys. {\bf 47} 042904 (2006)}\hspace{1cm}\footnotemark[13]\footnotetext{BCR}{S. Benenti,C. Chanu and G. Rastelli, J. Math. Phys. {\bf 41}, 4654 (2000)}
%\footnotemark[14]\footnotetext{RS}{P.Tempesta et al. eds: {\it Superintegrability in Classical and Quantum Systems}, CRM proceedings \& Lecture Notes {\bf 37} Amer. Math. Soc. (2004)} 

\subsection{Superintegrable 3-body systems on the line }  (see \footnote{C. Chanu, L. Degiovanni and G. Rastelli: {\it Superintegrable three body systems on the line} JMP 49, 112901 (2008)
}) By using Cylindrical coordinates $(r,\psi,z)$, with rotational axis $z$, and by indicating with $(p_r,p_\psi,p_z)$ their conjugate momenta, let us consider the natural Hamiltonian with potential 
\begin{equation}\label{V}
V=\frac{F(\psi)}{r^2}.
\end{equation}  
The potential $V$  is separable, and therefore Liouville integrable, in Cylindrical, Spherical, Parabolic, Ellipsoidal Prolate and Oblate coordinate systems. The five quadratic first integrals associated with the multiseparability take the form 
$H= \frac{1}{2} \left(p_r^2+\frac 1{r^2}p^2_\psi+p^2_z\right)+\frac {F(\psi)}{r^2}$, $H_1=\frac{1}{2}p_\psi^2+F(\psi)$, $H_2=\frac{1}{2}p_z^2$, $H_3=\frac{1}{2}\left[(rp_z-zp_r)^2+\left(1+\frac{z^2}{r^2}\right)p_{\psi}^2\right] +\left(1+\frac{z^2}{r^2}\right)F(\psi)$, $H_4=\frac{1}{2} \left(zp_r^2 +\frac{z}{r^2}p_{\psi}^2-rp_rp_z\right) + \frac{z}{r^2}F(\psi)$.
Only four of them are f.i.
The first three integrals allow the separation of the system in cylindrical coordinates. Let $x^i$ $i=1\ldots 3$ the positions of three points on a straight line, $(x^i)$ can be interpreted as Cartesian coordinates of a single point in the space. It is possible to show that
all interactions among the points of the form
$$
V=\sum _i\frac 1{X_i^2}F_i\left(\frac {X_{i+1}}{X_i},\frac{X_{i+2}}{X_i}\right) \quad i=1, \ldots, 3 \mbox{ (mod }3),
$$
are superintegrable with 4 f.i. first integrals, where $F_i$ are arbitrary  functions of two variables and $X_i=x^{i}-x^{i+1}, \quad i=1,2,3 \quad \mbox{(mod $3$)}$, $r\cos \psi =\frac1{\sqrt{2}}(x^1-x^2)$, $r\sin \psi =\frac1{\sqrt{6}}(x^1+x^2-2x^3)$, $z=\frac 1{\sqrt{3}}(x^1+x^2+x^3)$. Indeed, it can be proved that such $V$ are in the form (\ref{V}), then,  admit all the first-integrals $H_i$ four of which are always functionally independent. 
 
Examples:

 The {\bf Calogero system} 
$$
V_C=\frac{k_1}{(x^1-x^2)^2}+\frac{k_2}{(x^2-x^3)^2}+\frac{k_3}{(x^3-x^1)^2}= 
\sum_{i=1}^3 \frac {k_i}{X_i^2},\qquad k_i\in \mathbb R.
$$
The {\bf Wolfes system}
$$
V_W=\frac{k_1}{(x^1+x^3-2x^2)^2}+\frac{k_2}{(x^2+x^1-2x^3)^2}+\frac{k_3}{(x^3+x^2-2x^1)^2}
$$
$$
=\sum_i \frac{1}{X_i^2}k_{i+1}\left(\frac{X_{i+1}}{X_i}- \frac{X_{i+2}}{X_i} \right)^{-2}, \quad i=1, \ldots, 3 \mbox{ (mod }3).
$$
And a new one
$$
V=\sum_{i=1}^{3}\frac{k_i}{X_i^2+X_{i+1}^2}=\sum_{i=1}^3\frac {k_i}{X_{i+2}^2}\left(\frac{X_i^2}{X_{i+2}^2}+\frac{X_{i+1}^2}{X_{i+2}^2}\right)^{-1} \mbox{ (mod }3).
$$

%\footnotemark[15]\footnotetext{CDR}{C.Chanu, L.Degiovanni and G.Rastelli, arXiv: 0802.1353 }

\section{St\"ackel approximation} Several attempts have been made to use the St\"ackel systems as suitable integrable Hamiltonians to perturbe in order to approximate given physical systems. For example \footnote{J. Binney, S.  Tremaine, {\it Galactic Dynamics}, Princeton University press (1994)} and  \footnote{P. T. De Zeeuw, D. Lynden-Bell, MNRAS {\bf 215} 713 (1985)} for the dynamics of stars in elliptical galaxies. There, for given separable systems and by using the separable coordinates themselves, the possible separable potentials (i.e. satisfying the condition  $\mathbf d(K\;d V)=0$) are analyzed with the purpose  to find the separable dynamics closest to the gravitational potentials of elliptical galaxies. The best fitting integrable Hamiltonian is then determined by using perturbative methods. 

%\footnotemark[16]\footnotetext{BT}{J. Binney, S.  Tremaine, {\it Galactic Dynamics}, Princeton University press (1994)}\hspace{1cm}\footnotemark[17]\footnotetext{Dz}{P. T. De Zeeuw, D. Lynden-Bell, MNRAS {\bf 215} 713 (1985)}

\subsection{Decomposition of a potential into integrable and perturbative terms } (see \footnote{{G.Rastelli \it Decomposition of scalar potentials of natural Hamiltonians into integrable and perturbative terms: a naive approach}
J. Phys.: Conf. Ser. 128 (2008) 012029 (12pp)
}) Given a natural (non integrable) Hamiltonian $H$ we search for the a  separable system $(H_i)$, $i=1\ldots n$ representing "the closest", among all separable systems, approximation to the dynamics determined by $H$. Due to the structure of the separable systems, the quantities $m_i=\{H,L_i\}=dL_i/dt$ are linear homogeneous polynomials in the momenta. Then, the scalars $\mu_i=g_{jl}m_i^jm_i^l$  provide informations on the time-variation of the functions $L_i$ along the integral curves of $H$. It is remarkable that $d\mu_i=d\;K_id\;V$. We can conjecture that the minimization of all these quantities, which are coordinate-independent, among all the St\"ackel systems in some neighborhood of the configuration manifold determines the separable system which provides the best approximation of the given (nonintegrable) Hamiltonian dynamics in the same neighborhood. 

Example: 

Let us consider the {\bf Quadrupole field} in two dimensions:
$H=\frac 12 (p_x^2+p_y^2)+\frac Gr+\frac D{r^3}\left(3\frac{x^2}{r^2}-1\right)$
where $G,D$ are real constants. By applying the criteria of above we determine the separable Hamiltonian system associated to elliptic coordinates with foci of cartesian coordinates, if $D/C>0$, $x^2=2D/G, y=0$ and with potential $W_e=\frac 2{v^2-u^2}\left( Gv-4\frac Dv\right)$, where $(u,v)$ are the same elliptic coordinates. The dynamics of this  system can be compared with the original one and with the system separable in Polar coordinates \footnotemark[18].

%\ 

%\centerline{
%\includegraphics*[width=   140pt, height=   100pt]{Q_V_3}
%\includegraphics*[width=   140pt, height=   100pt]{Q_V_4}
%}
%\centerline{
%\includegraphics*[width=   140pt, height=   100pt]{Q_V_5}
%\includegraphics*[width=   140pt, height=   100pt]{Q_V_6}
%}
%\centerline{
%\includegraphics*[width=   140pt, height=   100pt]{Q_V_7}
%\includegraphics*[width=   140pt, height=   100pt]{Q_V_8}
%}
%\centerline{
%\includegraphics*[width=   140pt, height=   100pt]{Q_V_9}
%\includegraphics*[width=   140pt, height=   100pt]{Q_V_10}
%}

%For the same systems, the projection of the integral curves on the conjugate plane %$(x,p_x)$ is plotted:

%\footnotemark[18]\footnotetext{Dec}{G. Rastelli, {\it Decomposition of scalar potentials of natural Hamiltonians into integrable and perturbative terms: a naive approach}, to appear in J. Phys. Conference series: Proceedings of QTS5, Valladolid (2007) }

\section{Et cetera} Several advances in GTSOV cannot take place in this short presentation. It follows a short list just to mention them and to provide some references. {\bf Conformal separation}\footnote{S. Benenti, C. Chanu and G. Rastelli, J. Math. Phys. {\bf 46} 042901 (2005)} \footnote{M. Chanachowicz, C. Chanu and  R. G. McLenaghan, J. Math. Phys. {\bf 49} 013511 (2008)}, {\bf St\"ackel transformation}\footnote{E. G. Kalnins, J. M. Kress and W. Miller  Jr, J. Math. Phys {\bf 46} 053509 (2005)} \footnote{A. Sergeev and M. B\l aszak, J. Phys. A: Math. Theor. {\bf 41} 105205 (2008)}, {\bf Classification of separable systems} in constant curvature manifolds (3-dim Minkowski\footnote{J. T. Horwood and R. G. McLenaghan, J. Math. Phys. {\bf 49} 023501 (2008)}, 2-dim De Sitter and anti De Sitter\footnote{J. F. Cari\~nena, M. F. Ra\~nada, M. Santander and T. Sanz-Gil, J. Nlin. Math. Phys. {\bf 12} 230 (2005)}).

%\footnotemark[19]\footnotetext{BCRFE}{S. Benenti, C. Chanu and G. Rastelli, J. Math. Phys. {\bf 46} 042901 (2005)}\hspace{1cm}\footnotemark[20]\footnotetext{Ray1}{M. Chanachowicz, C. Chanu and  R. G. McLenaghan, J. Math. Phys. {\bf 49} 013511 (2008)}\hspace{1cm}\footnotemark[21]\footnotetext{KMK1}{E. G. Kalnins, J. M. Kress and W. Miller  Jr, J. Math. Phys {\bf 46} 053509 (2005)}\hspace{1cm}\footnotemark[22]\footnotetext{Ser}{A. Sergeev and M. B\l aszak, J. Phys. A: Math. Theor. {\bf 41} 105205 (2008)}\hspace{1cm}\footnotemark[23]\footnotetext{Ray2}{J. T. Horwood and R. G. McLenaghan, J. Math. Phys. {\bf 49} 023501 (2008)}\hspace{1cm}\footnotemark[24]\footnotetext{RS}{J. F. Carinena, M. F. Ranada, M. Santander and T. Sanz-Gil, J. Nlin. Math. Phys. {\bf 12} 230 (2005)}\hspace{1cm}

%\begin{textblock}{10}(5,23)
%\small{
%\textcolor{ssec2}{Conformal separation}
%Conformal sep 
%}
%\end{textblock}

% If you want to add a figure do something like this:

%\begin{textblock}{3}(0,15)
%  \begin{center}
%\resizebox{3\TPHorizModule}{!}{\includegraphics{my_figure.eps}}
%\\Figure 5: Googles per Snark (renormalised with wild angry men
%  \end{center}
%\end{textblock}

\end{document}